\documentclass[
a4paper,
prb,
aps,
reprint,
floatfix,
superscriptaddress,
]{revtex4-1}

\usepackage{graphicx}
\usepackage{amsmath}
\usepackage[colorlinks=true]{hyperref}

\begin{document}

\date{13 March 2017}

\title{Artificial control of the bias-voltage dependence of tunnelling
anisotropic magnetoresistance using quantization in a single-crystal
ferromagnet}

\author{Iriya Muneta}
\email{muneta.i.aa@m.titech.ac.jp}
\altaffiliation[Present address: ]{Department of Electrical and
Electronic Engineering, School of Engineering, Tokyo Institute of Technology,
4259 Nagatsuta-cho, Midori-ku, Yokohama, Kanagawa 226-8502, Japan.}
\affiliation{Department of Electrical Engineering and Information Systems, The
University of Tokyo, 7-3-1 Hongo, Bunkyo-ku, Tokyo 113-8656, Japan}

\author{Toshiki Kanaki}
\affiliation{Department of Electrical Engineering and Information Systems, The
University of Tokyo, 7-3-1 Hongo, Bunkyo-ku, Tokyo 113-8656, Japan}

\author{Shinobu Ohya}
\email{ohya@cryst.t.u-tokyo.ac.jp}
\affiliation{Department of Electrical Engineering and Information Systems, The
University of Tokyo, 7-3-1 Hongo, Bunkyo-ku, Tokyo 113-8656, Japan}
\affiliation{Center for Spintronics Research Network (CSRN), The University of
Tokyo, 7-3-1 Hongo, Bunkyo-ku, Tokyo 113-8656, Japan}

\author{Masaaki Tanaka}
\email{masaaki@ee.t.u-tokyo.ac.jp}
\affiliation{Department of Electrical Engineering and Information Systems, The
University of Tokyo, 7-3-1 Hongo, Bunkyo-ku, Tokyo 113-8656, Japan}
\affiliation{Center for Spintronics Research Network (CSRN), The University of
Tokyo, 7-3-1 Hongo, Bunkyo-ku, Tokyo 113-8656, Japan}

\begin{abstract}
A major issue in the development of spintronic memory devices is the reduction
of the power consumption for the magnetization reversal.
For this purpose, the artificial control of the magnetic anisotropy of
ferromagnetic materials is of great importance.
Here, we demonstrate the control of the carrier-energy dependence of the
magnetic anisotropy of the density of states (DOS)
using the quantum size effect in a
single-crystal ferromagnetic material, GaMnAs.
We show that the mainly two-fold symmetry of the magnetic anisotropy of DOS,
which is attributed to the impurity band, is changed to a four-fold symmetry
by enhancing the quantum size effect in the valence band of the GaMnAs quantum
wells.
By combination with the gate-electric field control technique,
our concept of the usage of the quantum size effect for the control of the
magnetism will pave the way for the ultra-low-power manipulation of
magnetization in future spintronic devices.
\end{abstract}

\maketitle

\begin{figure}
\centering
\includegraphics{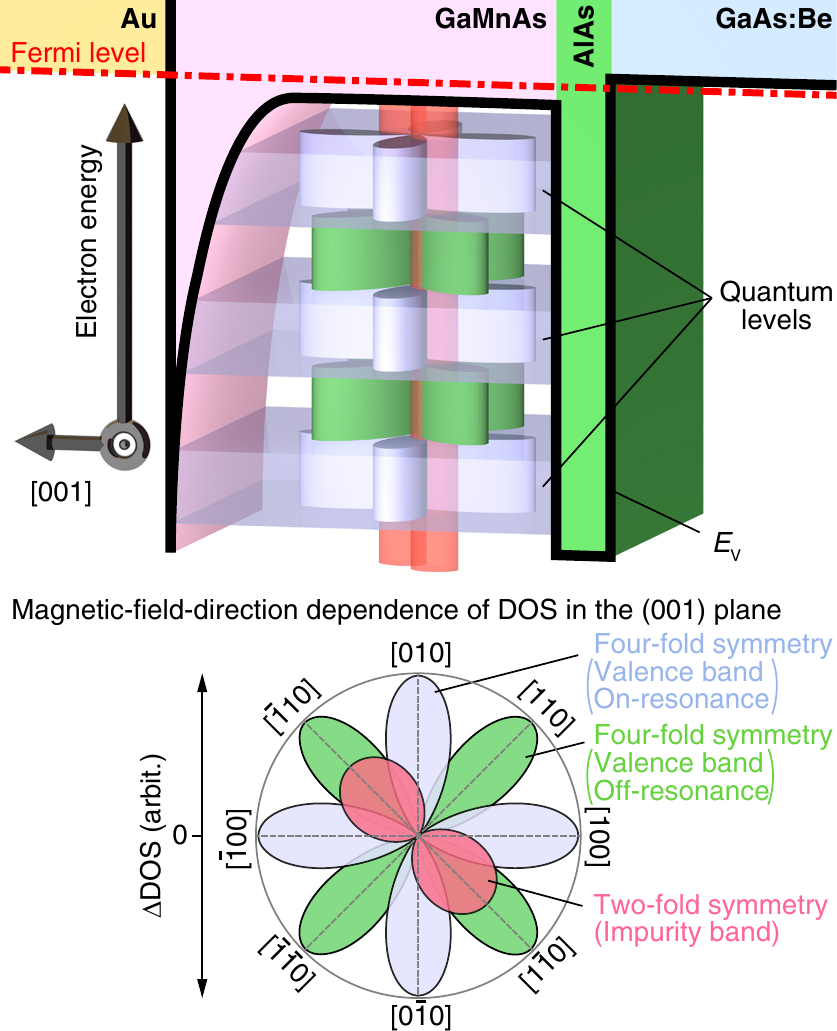}
\caption{\label{fig:front} \footnotesize
\noindent\textbf{Schematic valence band diagram representing our findings;
the mainly two-fold symmetry of the magnetic anisotropy of the density of
states (DOS),
which is attributed to the impurity band,
is changed to a four-fold symmetry by enhancing the quantum size effect in the
valence band in GaMnAs.}
The sample structure examined here is an Au / GaMnAs (22~nm, 16~nm and 9~nm)
quantum well / AlAs (5~nm) / $p$-GaAs:Be (100~nm) / $p^+$-GaAs (001) substrate,
from the top surface to the bottom.
The yellow, pink, green and blue regions
correspond to the Au, GaMnAs, AlAs and GaAs:Be layers, respectively.
The red dash-dotted line and black solid curve represent the Fermi level and
top energy of the valence band of each layer, respectively.
Three blue horizontal plates represent the quantum levels of the GaMnAs valence
band.
The two-fold symmetry (red) and four-fold symmetry (light blue and green)
represent the polar plots of the magnetic field direction dependence of the
$\Delta$DOS of the impurity band and quantized valence band, respectively.
Here, $\Delta$DOS is the change in the DOS and the azimuth is the in-plane
magnetization direction angle $\varphi$ of the GaMnAs layer.
These are based on the polar plots of the $dI/dV$-$\varphi$ curves shown in Fig.
\ref{fig:results-color-coded}c,i.
At the on-resonant states (blue plates),
the phase of the four-fold symmetry (light blue) is opposite to that at the
off-resonant states (green).
The two-fold symmetry,
which dominates the entire energy region,
is changed to four-fold symmetry as the
GaMnAs quantum well thickness is decreased and the quantum size effect is
enhanced.
}
\end{figure}

The control of the magnetic anisotropy \cite{Weisheit2007, Chiba2008,
Maruyama2009, Nozaki2010, Shiota2012, Nozaki2013, Rajanikanth2013} is a
particularly crucial issue for the reduction of the power required for the
magnetization reversal, which will enable the exploitation of
the full potential of the spintronic memory and logic devices, such as magnetic
tunnel junctions \cite{Miyazaki1995, Moodera1995, Tanaka2001, Yuasa2004,
Ikeda2010} and spin transistors \cite{Datta1990, Sugahara2004},
which may outperform the CMOS circuits commonly used in existing computers.
Meanwhile, band engineering is a useful technique to manipulate the electronic
structures of materials and heterostructures using an electric field
\cite{Bardeen1948}, the quantum size effect \cite{Esaki1970, Tsu1973}, and
modulation doping \cite{Dingle1978, Mimura1980},
and it is a well-established and important technology for the design of
electronic devices based on semiconductors.
For example, by designing quantum-well structures,
we can completely control the quantized energy levels or two-dimensional
sub-band structures of materials as calculated using band engineering.
Although band engineering was developed mainly for semiconductor
electronics,
it could potentially be extended to magnetism
because electron spins and their motions are coupled by spin-orbit interactions.
However, this point of view of the usage of band engineering is lacking
\cite{Ohno2000, Chiba2006, Sawicki2009, Xiu2010, Chiba2011, Shimamura2012,
Weisheit2007, Chiba2008, Maruyama2009, Nozaki2010, Shiota2012, Nozaki2013,
Rajanikanth2013}, and there are a wide variety
of possibilities for band engineering to design or control magnetic materials
and devices.
Here,
we demonstrate the artificial control of the carrier energy dependence of the
magnetic anisotropy of the density of states (DOS) for the first time
by designing quantum well
structures consisting of a single-crystal ferromagnetic thin film
and semiconductor barriers and tuning
the strength of the quantum size effect.
We used the prototypical ferromagnetic semiconductor GaMnAs,
which has a band structure in which the energy regions of
the valence band (VB) and impurity band (IB) overlap \cite{Kobayashi2014PRB}.
We found that the magnetic anisotropy of the DOS varies depending on the energy
of the carriers, which has not been reported in any ferromagnetic materials.
Furthermore, we revealed that the relative strength of the magnetic
anisotropy between the VB and IB,
which have four-fold and two-fold symmetries in the film plane, respectively,
can be varied by controlling the strength of the quantum size effect induced in
the VB (Fig. \ref{fig:front}).
These new findings suggest that band engineering provides the possibility of
artificially designing magnetic anisotropy by controlling the electronic
structures of magnetic materials.
By combining this technique with the recently developed gate electric-field
control technique of the carrier density and magnetization,
our results will lead to ultra-low-power manipulation of magnetization in future
spintronic devices.

\section*{Results}
\begin{figure}
\includegraphics{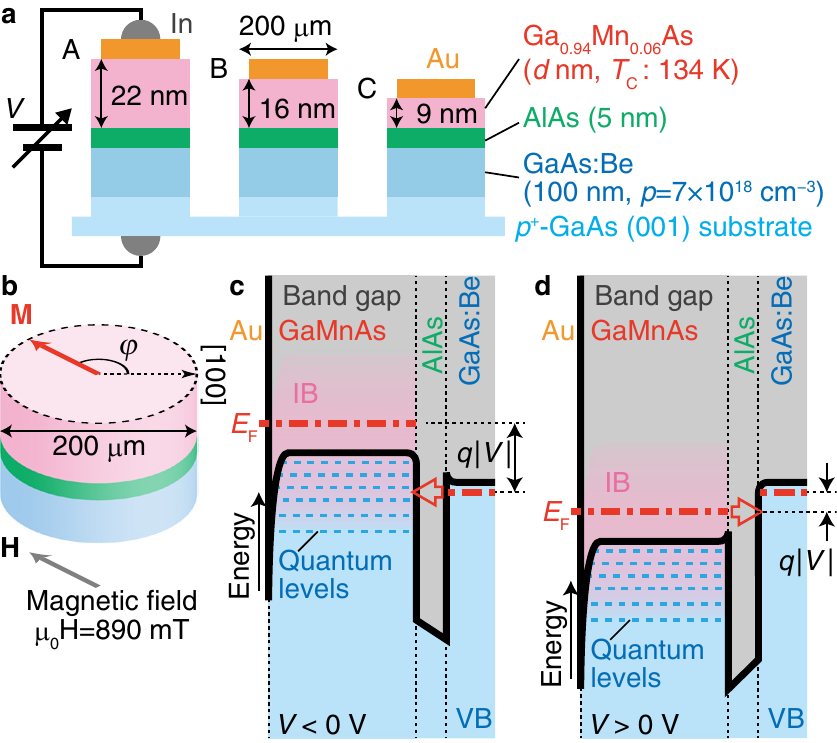}
\centering
\caption{\label{fig:structure}
\textbf{Device structure, magnetic field direction in our measurements and
valence band diagrams of our devices.}
  \textbf{a},
  Schematic cross-sectional structure of devices A-C used in this study.
  The bias voltage $V$ is applied between the Au electrode and backside of
  the substrate.
  \textbf{b},
  Schematic view representing the directions of the magnetization \textbf{M}
  (red arrow) of the GaMnAs layer and of the magnetic field
  $\mu_0|\mathbf{H}|$=890~mT (grey arrow) applied in our measurements (see
  Supplementary Note 1 and Supplementary Figure 1 for the relation between the
  directions of \textbf{H} and \textbf{M}).
  \textbf{c},\textbf{d},
  Schematic valence band (VB) diagrams of the tunnel devices when negative
  (\textbf{c}) and positive (\textbf{d}) $V$ are applied.
  The black solid and red dash-dotted lines represent the top of
  the VB and the quasi Fermi levels $E_\mathrm{F}$, respectively.
  The blue dash lines represent the quantized VB levels in the GaMnAs layer.
  These quantum levels are formed because the VB holes
  are confined by the surface Schottky barrier ($\sim$1~nm) and
  AlAs layer \cite{Ohya2011, Ohya2012}.
  The grey, blue and pink regions represent the band gap, VB and
  impurity band (IB), respectively.
  The red arrow represents the tunnel direction of the holes
  when $V$ is applied.
  The character $q$ represents the elementary charge.
}
\end{figure}

\noindent\textbf{Samples.}
We use a tunnel heterostructure consisting of Ga$_{0.94}$Mn$_{0.06}$As (25~nm,
Curie temperature $T_\mathrm{C}$: 134~K) / AlAs (5~nm) / GaAs:Be (100~nm,
hole concentration $p$=7$\times$10$^{18}$~cm$^{-3}$) grown on a $p^+$-GaAs (001)
substrate by molecular beam epitaxy (MBE) (see Fig. \ref{fig:structure}a).
We carefully etch a part of the GaMnAs layer from the surface and
fabricate three tunnel-diode devices with a diameter of 200 $\mu$m on the same
wafer \cite{Ohya2011, Ohya2012} with different GaMnAs thicknesses, $d$:
22~nm (device A), 16~nm (device B) and 9~nm (device C).
In the GaMnAs layer, holes are confined by the AlAs barrier and the thin
depletion layer ($\sim$1~nm) formed at the surface of the GaMnAs layer
\cite{Ohya2011, Ohya2012}, and thus, the VB energy levels in the GaMnAs layer
are quantized.
As shown in Fig. \ref{fig:structure}b,
we measure the current-voltage ($I$-$V$) characteristics at 3.5~K
with a strong magnetic field $\mu_0|\mathbf{H}|$=890~mT applied at an
angle $\varphi$ from the $[100]$ direction in the plane so that the
magnetization \textbf{M} becomes parallel to \textbf{H},
where $\mu_0$ is the vacuum permeability (see Supplementary Note 1 and
Supplementary Figure 1 for the relation between the directions of \textbf{H} and
\textbf{M}).
We ground the backside of the substrate
and apply a bias voltage $V$ to the Au top electrode.
In this manner, we measure the magnetization-direction dependence of the tunnel
conductance at various values of $V$ \cite{Gould2004, Saito2005, Ciorga2007,
Gao2007}.
For these measurements, we vary $V$ and $\varphi$ at
intervals of 2~mV and $5^\circ$, respectively.
As shown in Fig. \ref{fig:structure}c,
when a negative $V$ is applied,
holes tunnel from the GaAs:Be layer to the GaMnAs layer.
Because $dI/dV$ is proportional to the DOS at the energy where tunnelling
occurs,
the energy dependence of the DOS below the Fermi level $E_\mathrm{F}$ of the
GaMnAs layer is detected in negative $V$.
When $V$ is positive,
holes tunnel from the GaMnAs layer to the GaAs:Be layer,
and thus, $dI/dV$ is proportional to the DOS at $E_\mathrm{F}$
of GaMnAs regardless of $V$ (see Fig. \ref{fig:structure}d).
Our measurements provide the $\varphi$ dependence of the DOS at various
energies below $E_\mathrm{F}$ in GaMnAs.

\begin{figure*}
\includegraphics{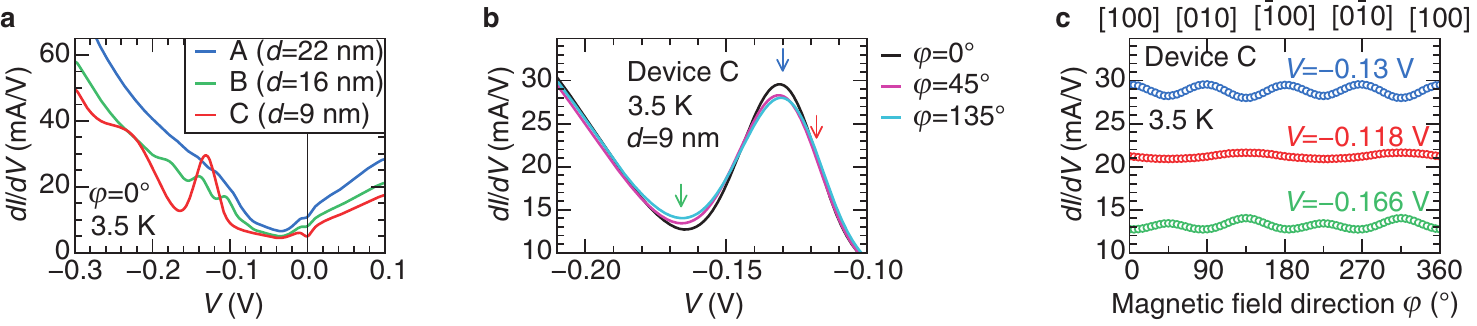}
\centering
\caption{\label{fig:results-simple}
\textbf{Measurement results of the tunnel transport in devices A-C.}
  \textbf{a},
  Comparison of $dI/dV$-$V$ curves
  measured on different devices when the magnetic field direction $\varphi$ is
  $0^\circ$.
  \textbf{b},
  Comparison of three $dI/dV$-$V$ characteristics at different $\varphi$.
  The arrows indicate the $V$ values used in \textbf{c}.
  \textbf{c},
  The $\varphi$ dependence of $dI/dV$ in device C when $V$ is fixed at $-0.13$
  V (blue), $-0.118$ V (red) and $-0.166$ V (green),
  which correspond to the arrows shown in \textbf{b}.
  }
\end{figure*}

\noindent\textbf{Analysis.}
The $dI/dV$-$V$ characteristics obtained for devices A-C at $\varphi=0^\circ$
show different oscillatory behaviour (see Fig. \ref{fig:results-simple}a);
the oscillation becomes stronger as $d$ decreases.
Also, the $V$ values of the $dI/dV$ peaks
systematically change by changing $d$,
indicating that
these oscillations originate from the resonant tunnelling effect
\cite{Esaki1966, Grobis2005} induced by the quantum size effect in the VB of the
GaMnAs layer \cite{Ohya2007, Ohya2010, Ohya2011, Muneta2013, Tanaka2014}.
[See Supplementary Figure 3 for the $I$-$V$ characteristics of devices A-C.
More systematic data of the $d^2I/dV^2$-$V$ characteristics of tunnelling
devices with various GaMnAs thicknesses ($d$ =  7.3~nm -- 23.6~nm) are described
in Supplementary Note 3 and Supplementary Figure 4.]
The oscillation amplitude of $dI/dV$ decreases by changing $\varphi$
from $0^\circ$ (along $[100]$) to $45^\circ$ (along $[110]$) or $135^\circ$
(along $[\bar{1}10]$) (see Fig. \ref{fig:results-simple}b).
The peak (blue arrow) value of $dI/dV$ when $\varphi = 0^\circ$ is larger than
that when $\varphi = 45^\circ$ or $135^\circ$,
whereas the dip (green arrow) value of $dI/dV$ when $\varphi = 0^\circ$ is
smaller than that when $\varphi = 45^\circ$ or $135^\circ$.
This feature can been seen in Fig. \ref{fig:results-simple}c as the opposite
sign of the oscillation of $dI/dV$ as a function of $\varphi$ between when $V =
-0.13~\mathrm{V}$ (the peak of $dI/dV$-$V$ in Fig. \ref{fig:results-simple}b)
and when $V = -0.166~\mathrm{V}$ (the dip of $dI/dV$-$V$ in Fig.
\ref{fig:results-simple}b).
In Fig. \ref{fig:results-simple}c,
the symmetry of the $\varphi$ dependence of $dI/dV$ changes depending on $V$.
\begin{figure*}
\includegraphics[width=\textwidth,bb=0.000000 0.297852 480.753000 169.214000]
{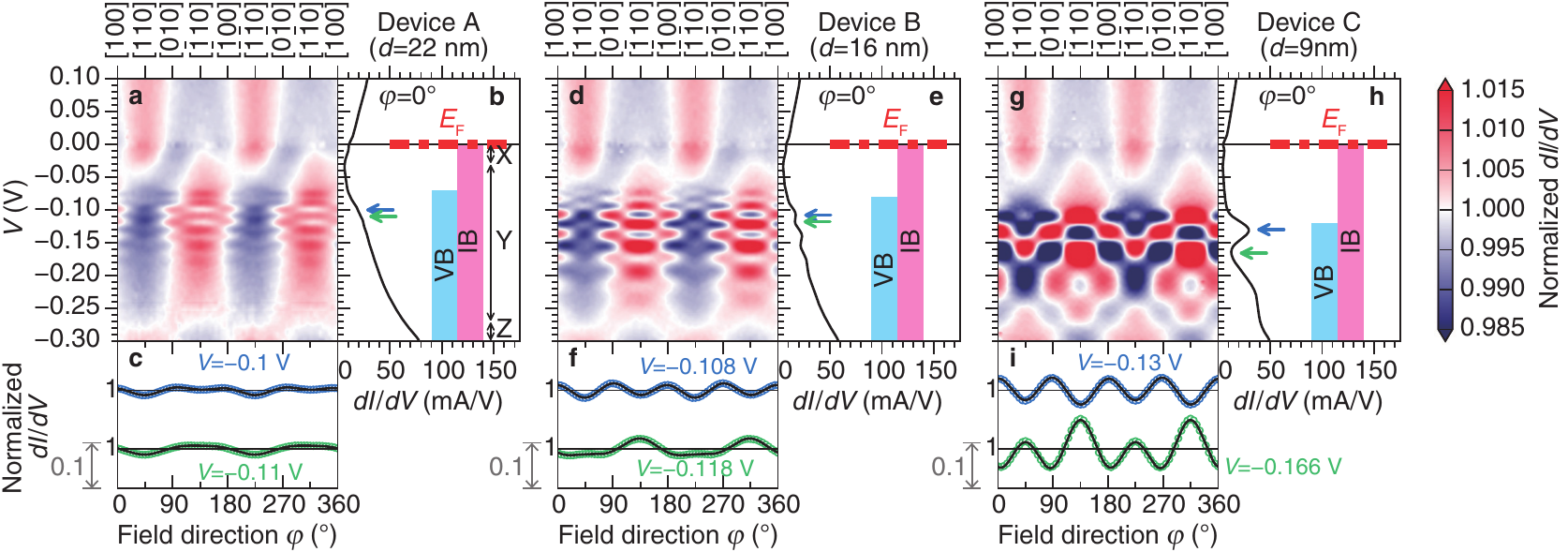}
\centering
\caption{\label{fig:results-color-coded}
\textbf{Measurement results of the magnetization-direction dependence of the
tunnel transport in devices A-C.}
  \textbf{a},\textbf{d},\textbf{g},
  Colour-coded maps representing the normalized $dI/dV$ as a function
  of the magnetic field direction $\varphi$ and $V$.
  The normalized $dI/dV$ is defined by Equation \eqref{eq:normalization}.
  The negative voltage region in device A is classified into three parts X, Y
  and Z by the sign of the oscillation in the normalized $dI/dV$-$\varphi$
  curves.
  \textbf{b},\textbf{e},\textbf{h},
  Characteristics of $dI/dV$-$V$ at $\varphi=0^\circ$.
  The blue and green arrows indicate $V$ corresponding to the peak and dip,
  respectively,
  which are used in \textbf{c}, \textbf{f} and \textbf{i}.
  The blue and pink regions shown on the right side represent the valence
  band (VB) and impurity band (IB) regions in GaMnAs, respectively.
  The red dash-dotted lines represent the Fermi level $E_\mathrm{F}$ ($V$=0).
  \textbf{c},\textbf{f},\textbf{i},
  The normalized $dI/dV$-$\varphi$ curves at $V$ corresponding to the positions
  of the blue and green arrows in \textbf{b}, \textbf{e} and \textbf{h},
  respectively.
  The black solid curves are the fitting curves expressed by Equation
  \eqref{eq:fitting}.
  }
\end{figure*}
Here, we normalize $dI/dV$ using
\begin{equation}
\label{eq:normalization}
\left( \mathrm{Normalized} \; dI/dV \right) =
\frac{dI/dV}{\langle dI/dV \rangle_\varphi},
\end{equation}
where $\langle dI/dV \rangle_\varphi$ is the $dI/dV$ value averaged over
$\varphi$ at a fixed $V$.
As seen in Fig. \ref{fig:results-color-coded}a,d,g,
primarily two-fold symmetry along $[110]$ is observed at negative $V$ when $d$
= 22 nm,
but four-fold symmetry along $\langle 100 \rangle$ emerges as $d$ decreases.
At the peaks and dips of the $dI/dV$-$V$ curves indicated by the blue and
green arrows in Fig. \ref{fig:results-color-coded}b,e,h,
the symmetry of the $dI/dV$-$\varphi$ curves shown in
Fig. \ref{fig:results-color-coded}c,f,i is changed from two-fold to four-fold as
the resonant tunnelling is enhanced by decreasing $d$.
Because resonant tunnelling is induced in the VB,
the enhanced four-fold symmetry is attributed to the VB.
The curves in Fig. \ref{fig:results-color-coded}i are those obtained by
normalizing the $dI/dV$-$\varphi$ curves shown in Fig.
\ref{fig:results-simple}c (blue and green).

\begin{figure*}
\centering
\includegraphics{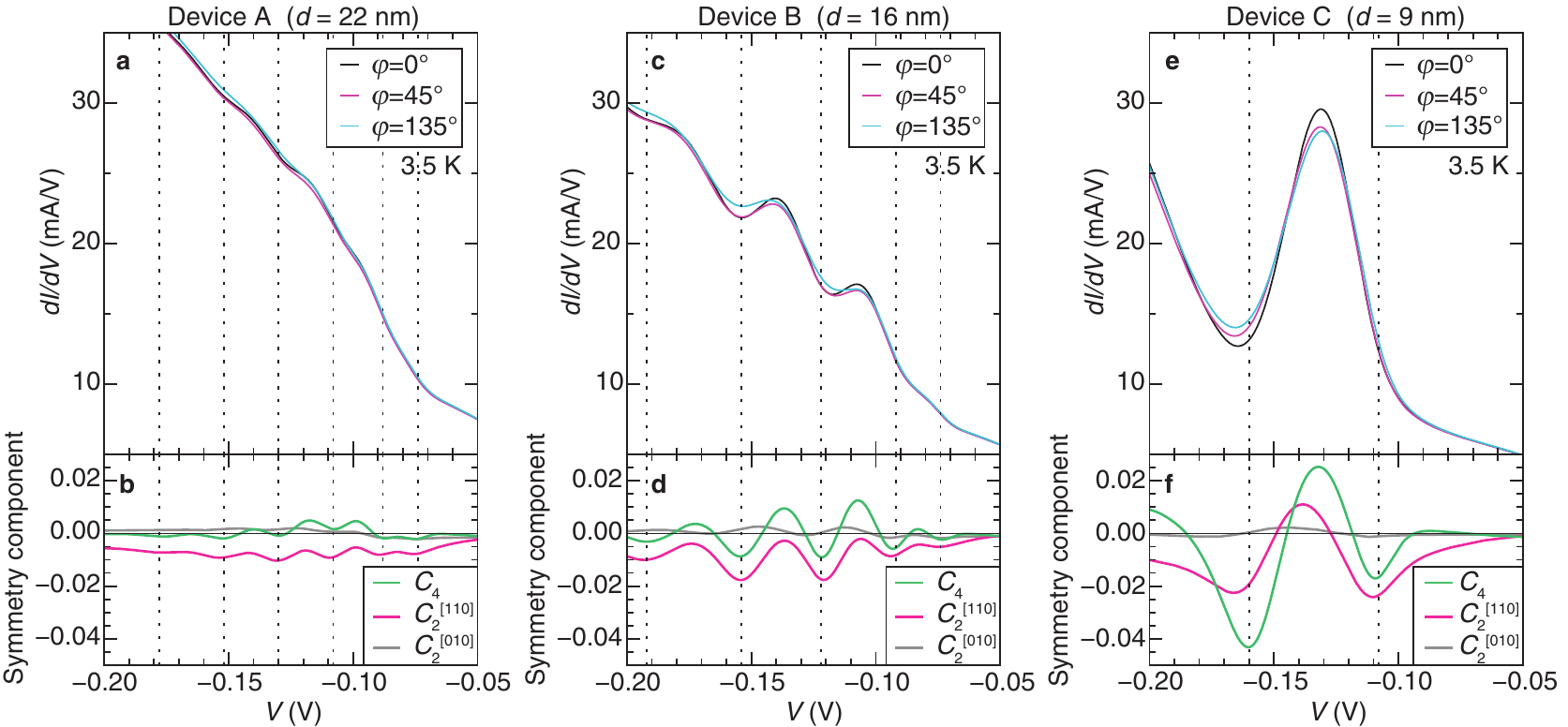}
\caption{\label{fig:sync}
\textbf{
Relationship between the {\boldmath $dI/dV$}-{\boldmath $V$}
characteristics and {\boldmath$V$} dependence of the symmetry components
{\boldmath $C_4$}, {\boldmath $C_2^{[110]}$} and {\boldmath $C_2^{[010]}$}.
}
\textbf{a},\textbf{c},\textbf{e},
The $dI/dV$-$V$ characteristics measured at $\varphi=0^\circ$ (black),
$\varphi=45^\circ$ (pink) and $\varphi=135^\circ$ (light blue) on devices A-C,
respectively.
\textbf{b},\textbf{d},\textbf{f},
The four-fold symmetry components $C_4$ along $\langle 100 \rangle$ (green),
two-fold symmetry components $C_2^{[110]}$ along $[110]$ (pink) and two-fold
symmetry components $C_2^{[010]}$ along $[010]$ (grey) as a function of $V$.
These components are obtained by fitting the curves expressed by equation
\eqref{eq:fitting} to the normalized $dI/dV$-$\varphi$ curves at each $V$.
The vertical dotted lines in the upper and lower panels represent the $V$ at
which $C_4$ reaches dips.
}
\end{figure*}

We derive the symmetry components
$C_4$ (four-fold along $\langle 100 \rangle$),
$C_2^{[110]}$ (two-fold along $[110]$) and
$C_2^{[010]}$ (two-fold along $[010]$)
of the normalized $dI/dV$-$\varphi$ curves
by fitting the following equation
to the experimental normalized $dI/dV$-$\varphi$ curves:
\begin{widetext}
\begin{equation}
\label{eq:fitting}
\left( \mathrm{Normalized} \; dI/dV \right) =
C_4 \cos 4\varphi
+ C_2^{[110]} \cos 2 \left( \varphi - 45^\circ \right)
+ C_2^{[010]} \cos 2 \left( \varphi - 90^\circ \right)
+ 1.
\end{equation}
\end{widetext}
The $V$ dependences of $C_4$, $C_2^{[110]}$ and $C_2^{[010]}$ show oscillations,
which synchronize with the oscillations of the $dI/dV$-$V$ curves that are
induced by the resonant tunnelling (see Fig. \ref{fig:sync}).
$C_4$ is significantly enhanced as $d$ is decreased and the oscillation of
$dI/dV$-$V$ is enhanced.
Therefore, the oscillatory behaviour of the $V$ dependence of the symmetry
components is attributed to the quantization of the VB states.

We discuss the origin of the four-fold symmetry of the
$dI/dV$-$\varphi$ curves induced by the quantum size
effect in the VB of GaMnAs.
Our results mean that the strength of the quantization depends on $\varphi$ as
shown in Fig. \ref{fig:results-simple}b, in other words,
the coherence length of the VB holes depends on $\varphi$.
In GaMnAs, there is a weak interaction between the VB holes and
Mn spin magnetic moments, which indicates that the coherence length of the
VB holes depends on the strength of this interaction.
Thus, the four-fold symmetry of the magnetic anisotropy
originates from the anisotropy of the wave function of the VB holes,
which are mainly composed of As $4p$ orbitals located at the lattice points
having a four-fold symmetry in the film plane.
This shows that the interaction between the spins and orbitals reflects
the anisotropy of the wave function distribution and the direction of spins.
We calculated the DOS of the VB, which is weakly interacting with the magnetic
moments of Mn, using the {\boldmath$k$}$\cdot${\boldmath$p$} Hamiltonian and
$p$-$d$ exchange Hamiltonian, and confirmed that the DOS \textit{vs.} $\varphi$
characteristic shows the four-fold symmetry (see Supplementary Note 4 and
Supplementary Figure 5).

\noindent\textbf{Magnetic anisotropy of the IB.}
In the region of $V$ = $-0.07$~V -- +0.1~V of the colour-coded maps shown in
Fig. \ref{fig:results-color-coded}a,d,g,
the normalized $dI/dV$ as a function of $\varphi$ and
$V$ is similar in devices A-C,
which means that it is insensitive to the change in $d$,
and thus the two-fold symmetry along $[110]$ observed in this region is
attributed to the IB.
This insensitivity to $d$ agrees with the previous report \cite{Proselkov2012}
that the magnetic anisotropy of magnetization in GaMnAs films at low
temperature is independent of $d$.
In the region of $V > 0$ V,
the $\varphi$ dependence of the normalized $dI/dV$ does not depend on $V$
because it always reflects the DOS at $E_\mathrm{F}$ in GaMnAs regardless of
$V$.
The two-fold symmetry along $[110]$ is also observed in the region of $V <
-0.07$~V for device A ($d$ = 22 nm, Fig. \ref{fig:results-color-coded}a),
in which the effect of the quantization of the VB holes is small.
This indicates that the $\varphi$ dependence of the tunnelling transport is
dominated by the IB holes in the entire region of $V$ when the quantization of
the VB holes is weak.
However, as the quantization becomes stronger,
the four-fold symmetry originating from the VB emerges in the IB region,
which means that the IB and VB overlap (Fig.
\ref{fig:results-color-coded}b,e,h).
This finding is consistent with recent angle-resolved photoemission
spectroscopy measurements of GaMnAs \cite{Kobayashi2014PRB}.

As shown in Fig. \ref{fig:results-color-coded}b, we can classify the region of
$V < 0$ V (corresponding to the DOS below $E_\mathrm{F}$ in GaMnAs) into three
parts, from top to bottom:
$0$~V -- $-0.03$~V (region X), $-0.03$~V -- $-0.27$~V (region Y) and $-0.27$~V
-- $-0.3$~V (region Z).
In regions X and Z, the DOS when \textbf{M} is along $[110]$ and
$[\bar{1}\bar{1}0]$ is larger than when \textbf{M} is along $[\bar{1}10]$
and $[1\bar{1}0]$,
whereas it is smaller in region Y.
We discuss the origin of this sign change of the $\varphi$ dependence of the DOS
of the IB depending on the energy.
A single Mn atom doped into GaAs forms an impurity state
because of the inter-atomic interaction (hybridization) between the Mn $3d$
orbitals and the As $4p$ orbitals.
Tang and Flatt\'e predicted that the hybridization and spin-orbit
interaction in the As $4p$ orbitals result in an anti-parallel condition
between the spin angular momentum of the Mn $3d$ spins and the orbital angular
momentum of the hole in the impurity state \cite{Tang2005}.
Because of this condition,
the wave function of the hole in the impurity state favours extension in the
direction perpendicular to the Mn $3d$ spins,
and thus the distribution of the wave function depends on the direction of the
Mn $3d$ spins.
This behaviour is well reproduced by a tight binding method (see Supplementary
Note 5 and Supplementary Figure 6).
\begin{figure}
\includegraphics{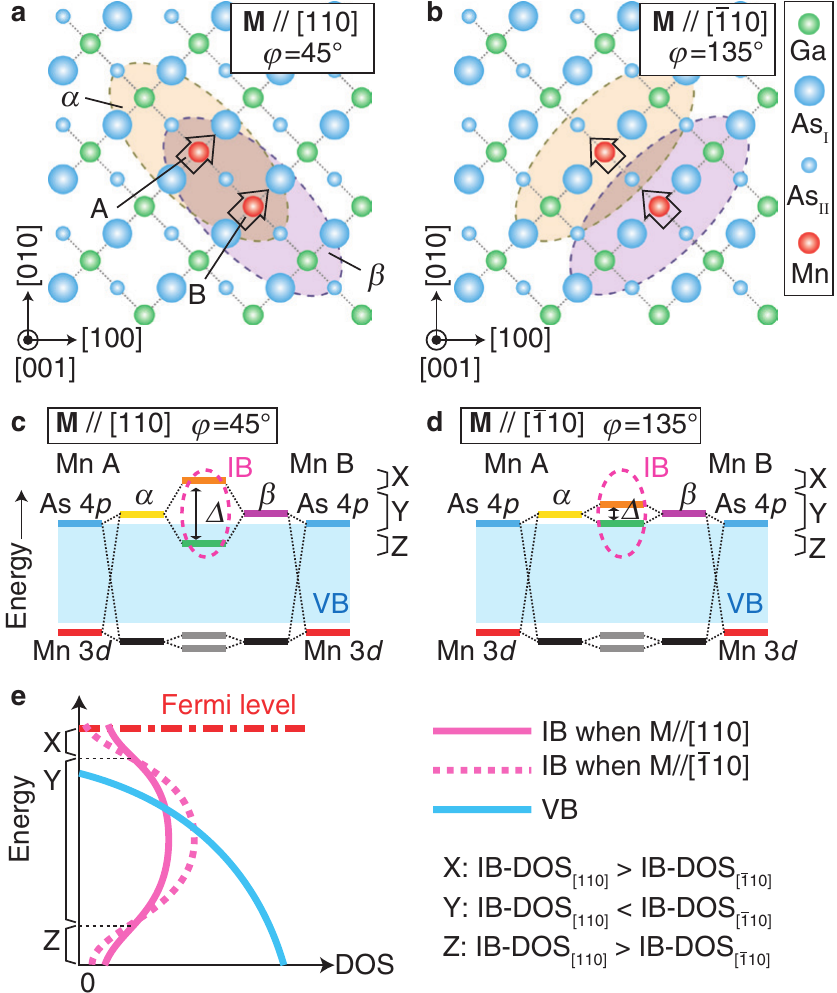}
\centering
\caption{\label{fig:cartoon}
\textbf{Diagrams of the wave functions, energy levels and densities of states
(DOSs) when the magnetization M direction {\boldmath $\varphi$} is along
the {\boldmath $[110]$} and {\boldmath $[\bar{1}10]$} axes in GaMnAs.}
  \textbf{a} and \textbf{b},
  Schematic three adjoining (001) planes of GaMnAs:
  the upper As (As$_\mathrm{I}$), middle Ga (or Mn) and lower As
  (As$_\mathrm{II}$) planes when \textbf{M}
  is along the $[110]$ (\textbf{a}) and $[\bar{1}10]$ (\textbf{b}) directions.
  The ellipses schematically represent the wave functions of
  the impurity states $\alpha$ (yellow) and $\beta$ (purple) originating from
  Mn A and B, respectively.
  The arrows represent the spin magnetic moments of Mn A and B.
  \textbf{c},\textbf{d},
  Schematic energy-level diagrams when \textbf{M} is along
  the $[110]$ (\textbf{c}) and $[\bar{1}10]$ (\textbf{d}) directions.
  In each graph, the left-most (right-most) lines correspond to the
  $3d$ level (red) of Mn A (B) and its neighbouring As $4p$ level (blue).
  The yellow and purple lines represent $\alpha$ and $\beta$, respectively.
  The wavefunction overlap between $\alpha$ and $\beta$ induces the bonding
  (green line) and anti-bonding (orange line) states with energy separation
  $\Delta$.
  The black lines represent the localized states 
  around Mn A and B formed by the hybridization between the $3d$ and $4p$
  orbitals.
  The wave functions of these states overlap slightly,
  which induces the anti-bonding and bonding states with small energy
  separation (grey lines).
  \textbf{e},
  Schematic DOS around the top of the valence band (VB) in GaMnAs.
  The pink solid and dotted curves represent the DOS of the impurity band (IB)
  when \textbf{M} is along $[110]$ and $[\bar{1}10]$, respectively.
  The blue solid curve represents the DOS of VB.
}
\end{figure}
Figure \ref{fig:cartoon}a,b
schematically illustrates the distributions of the wave functions of
the two impurity states $\alpha$ and $\beta$ originating from the two
neighbouring Mn atoms A and B located along $[\bar{1}10]$, respectively.
Here, we consider the cases in which
\textbf{M} is aligned in the $[110]$ direction (Fig. \ref{fig:cartoon}a) and in
the $[\bar{1}10]$ direction (Fig. \ref{fig:cartoon}b), in which the wave
functions tend to extend in the $[\bar{1}10]$ and $[110]$ directions,
respectively (the calculated results are shown in Supplementary Figure 6).
As shown in Fig. \ref{fig:cartoon}c,d,
the hybridization between the Mn $3d$ (red lines) and As $4p$
(blue lines) orbitals forms $\alpha$ (yellow line) and $\beta$ (purple line)
around Mn atoms A and B, respectively.
The bonding (green line) and anti-bonding (orange line) impurity states are
formed
by the wavefunction overlap between $\alpha$ and $\beta$,
which is the origin of the IB \cite{Mahadevan2004PRL, Kitchen2006}.
As indicated in Fig. \ref{fig:cartoon}a,b,
the wavefunction overlap between $\alpha$ and $\beta$ is larger when \textbf{M}
$\parallel$ $[110]$ than that when \textbf{M} $\parallel$ $[\bar{1}10]$.
Thus,
the energy separation $\Delta$ between the bonding and anti-bonding states
when \textbf{M} $\parallel$ $[110]$ is
larger (Fig. \ref{fig:cartoon}c) than that when \textbf{M} $\parallel$
$[\bar{1}10]$ (Fig. \ref{fig:cartoon}d).
The energy regions X-Z indicated in Fig. \ref{fig:cartoon}c,d correspond to the
regions of $V$ shown in Fig. \ref{fig:results-color-coded}b.
When \textbf{M} $\parallel$ $[110]$,
the anti-bonding and bonding states are formed in regions
X and Z, respectively (Fig. \ref{fig:cartoon}c).
When \textbf{M} $\parallel$ $[\bar{1}10]$,
both anti-bonding and bonding states are formed in region
Y (Fig. \ref{fig:cartoon}d).
Thus, the energy dependence of the DOS of the IB differs depending on the
\textbf{M} direction (i.e., $[110]$ or $[\bar{1}10]$), as shown in Fig.
\ref{fig:cartoon}e.
In regions X and Z, the DOS when
\textbf{M} $\parallel$ $[110]$ is larger than that when
\textbf{M} $\parallel$ $[\bar{1}10]$.
In region Y, the DOS when \textbf{M} $\parallel$ $[\bar{1}10]$ is larger
than that when \textbf{M} $\parallel$ $[110]$.
The same scenario can be applied for the Mn atoms whose distance is larger than
that between the nearest Mn atoms.
The two-fold symmetry along the $[110]$ axis in the DOS \textit{vs.}
$\varphi$ characteristic is well reproduced by a tight-binding calculation (see
Supplementary Note 5 and Supplementary Figure 7).

Above, we considered only the case in which the Mn atoms are located along
$[\bar{1}10]$.
Although there are several possible Mn alignments in real GaMnAs,
there are reasons that only the interaction between the Mn atoms located
along $[\bar{1}10]$ is important.
The overlap between the wave functions of the impurity states originating from
the two Mn atoms located along $\langle 110 \rangle$ is known to be larger
than that from the Mn atoms along other directions, such as  $\langle 100
\rangle$ and $\langle 211 \rangle$ \cite{Mahadevan2004PRL, Kitchen2006}.
Furthermore, an anisotropic distribution of Mn
atoms along $[\bar{1}10]$ in GaMnAs is predicted \cite{Birowska2012}.
Slightly more Mn atoms are located along $[\bar{1}10]$ than along $[110]$,
which is attributed to the direction of the Mn-As bonds on the surface during
the MBE growth.
This is thought to be the origin of the two-fold symmetry along $[110]$ of the
$dI/dV$-$\varphi$ curves.
Also, the anisotropic interaction between two Mn atoms and the
anisotropic distribution of the Mn atoms are thought to be the reason that
GaMnAs has magnetic anisotropy with
the in-plane two-fold symmetry along $[\bar{1}10]$.

$C_2^{[110]}$ is also slightly enhanced at the peaks and dips of the resonant
oscillation in $dI/dV$-$V$ as $d$ decreases (see Fig. \ref{fig:sync}),
which indicates that the VB has a small two-fold symmetry along $[110]$.
The two-fold symmetry of the VB is induced because
the interaction between the VB holes and Mn spin magnetic moments
transmits the anisotropy of the Mn distribution to the VB.

\begin{figure*}
\centering
\includegraphics{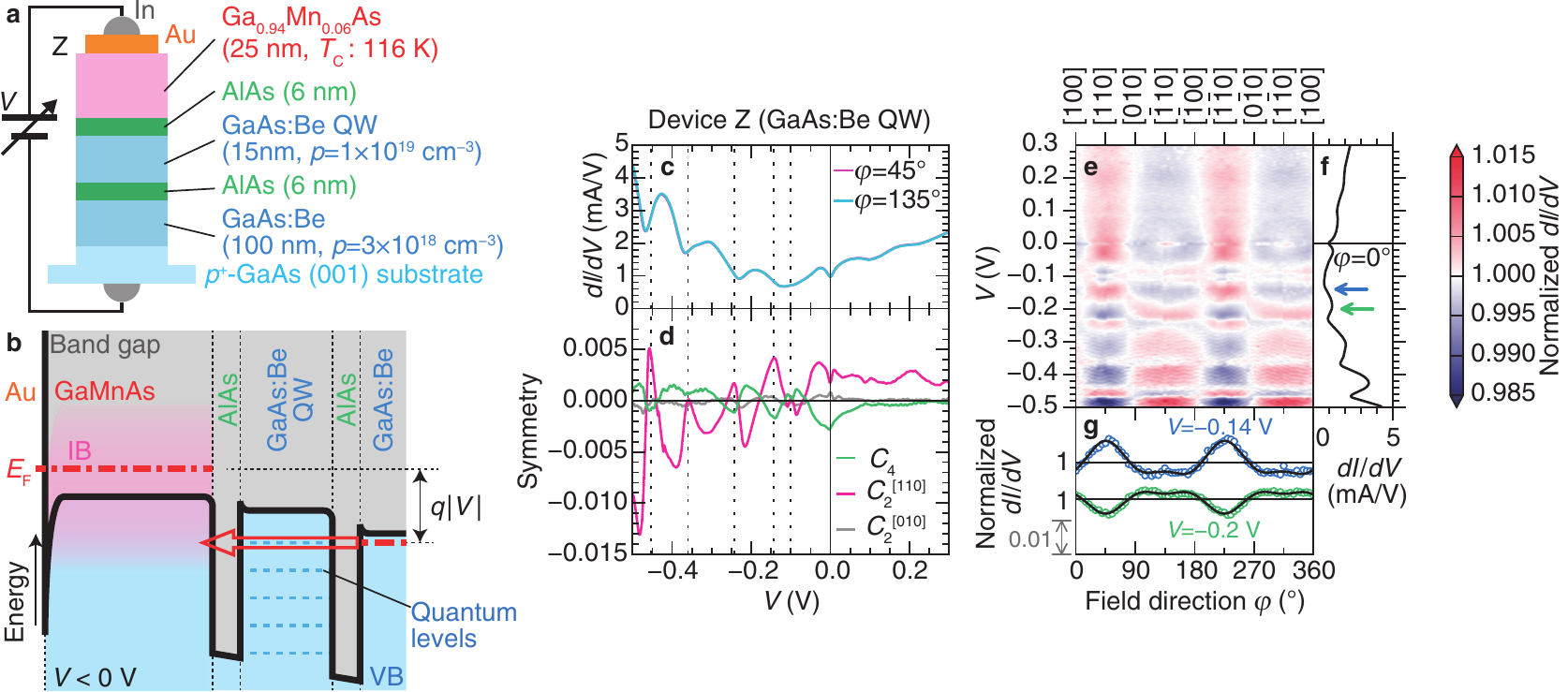}
\caption{\label{fig:comparison}
\textbf{Measurement results on device Z, with a non-magnetic GaAs:Be quantum
well (QW) and a ferromagnetic GaMnAs electrode for comparison.
}
  \textbf{a},
  Schematic cross-sectional structure of device Z used in this study.
  The bias voltage $V$ is applied between an Au electrode and the backside of
  the substrate.
\textbf{b},
  Schematic valence-band (VB) diagram of device Z when a negative $V$ is
  applied. See Fig. \ref{fig:structure}c for the legend.
\textbf{c},
  Obtained $dI/dV$-$V$ characteristics with the magnetic field directions
  $\varphi$ = $45^\circ$ (pink) and $135^\circ$ (light blue) at 3.5~K.
  These two curves are nearly completely overlapping.
\textbf{d},
  The four-fold symmetry component along $C_4$ $\langle 100 \rangle$ (green),
  two-fold symmetry component $C_2^{[110]}$ along $[110]$ (pink) and two-fold
  symmetry component $C_2^{[010]}$ along $[010]$ (grey) as a function of $V$.
  These components are obtained by fitting the curves expressed by Equation
  \eqref{eq:fitting} to the normalized $dI/dV$-$\varphi$ curves at each $V$.
  The vertical dotted lines in \textbf{c} and \textbf{d} represent the $V$ at
  which $C_4$ reaches a dip.
\textbf{e},
  Colour-coded map representing the normalized $dI/dV$ as a function of
  $\varphi$ and $V$.
\textbf{f},
  Characteristic of $dI/dV$-$V$ at $\varphi=0^\circ$.
  The blue and green arrows indicate the $V$ of the dip and peak,
  which are used in \textbf{g}.
\textbf{g},
  Normalized $dI/dV$-$\varphi$ curves at $V=-0.14$ V (blue) and $V=-0.2$ V
  (green) indicated by the blue and green arrows in \textbf{f}, respectively.
  The black solid curves are obtained by the fitting.
}
\end{figure*}

\noindent\textbf{Comparison.}
To verify that the four-fold symmetry of the magnetic anisotropy of DOS is
induced by the quantization in the GaMnAs quantum well (QW),
we compare devices A-C with device Z, which consists of GaMnAs (25~nm,
$T_\mathrm{C}$ 116~K) / AlAs (6~nm) / GaAs:Be QW (15~nm) / AlAs (6~nm) /
GaAs:Be (100~nm) grown on a $p^+$-GaAs (001) substrate (see Fig.
\ref{fig:comparison}a,b), in which the ferromagnetic layer and the QW layer are
separated and the quantum size effect does not occur in the ferromagnetic
layer.
The surface GaMnAs layer is thick enough to prevent the quantum size effect in
this layer.
We perform the same measurement on device Z.
The obtained $dI/dV$-$V$ characteristics oscillate because of the
quantum size effect in the nonmagnetic GaAs:Be QW (see Fig.
\ref{fig:comparison}c).
The $V$ dependence of the normalized $dI/dV$-$\varphi$ curves reflects the
energy dependence of the magnetic anisotropy of the DOS of the GaMnAs electrode
and exhibits an oscillatory behaviour, which is attributed to the oscillation
of the $dI/dV$-$V$ curves induced by the resonant tunnelling effect in the
GaAs:Be QW (see Fig. \ref{fig:comparison}c-f).
Similarly to devices A-C,
$C_4$, $C_2^{[110]}$ and $C_2^{[010]}$ oscillate as a function of $V$,
synchronizing with the oscillation of $dI/dV$-$V$
(Fig. \ref{fig:comparison}c,d);
however, the symmetry of the $\varphi$ dependence of the normalized $dI/dV$ is
mainly two-fold, reflecting the magnetic anisotropy of the GaMnAs top electrode
(see Fig. \ref{fig:comparison}d,e,g),
which offers a remarkable contrast to the results of devices A-C that the
quantum size effect in the GaMnAs QW layer enhances $C_4$ (four-fold symmetry).
This contrast provides evidence that
the magnetic anisotropy of the DOS changes by enhancing the quantum size
effect in the GaMnAs QW in devices A-C.
Between the different $V$ values indicated by the green and blue arrows in
Fig. \ref{fig:comparison}f, the symmetry of the $dI/dV$-$\varphi$
curves shows an opposite sign (Fig. \ref{fig:comparison}g),
which is induced by the small shift of the peak $V$ of the $dI/dV$-$V$ curves
(see Supplementary Note 2 and Supplementary Figure 2 for details).

It should be noted that surface states, if any, may have different anisotropy
from the bulk or quantum well states.
However, the surface of GaMnAs is depleted and does not induce the
carrier-mediated ferromagnetism.
This means that the surface does not couple with the magnetization of the
GaMnAs layer beneath the surface.
Thus, the DOS of the non-magnetic surface does not depend on the direction of
the magnetization of the GaMnAs layer.

The magnetic anisotropy of magnetization in ferromagnetic materials reflects
that of the DOS at $E_\mathrm{F}$ \cite{Saito2005}.
Our study indicates that
the magnetic anisotropy of DOS depends on the carrier energy (applied voltage)
and can be controlled by band engineering.
Combining our results with the electric-field gating technique to tune the
$E_\mathrm{F}$ position
will provide a new method to manipulate the magnetization direction by
controlling the magnetic anisotropy with an ultra-low power.
This method will be useful for the development of non-volatile spin devices
using magnetization in the future.

\section*{Methods}
\noindent\textbf{Sample preparation.}
We grew a Ga$_{0.94}$Mn$_{0.06}$As (25~nm) / AlAs (5~nm) / GaAs:Be (100~nm,
$p=7\times10^{18}$ cm$^{-3}$) tunnel heterostructure by MBE for the fabrication
of devices A-C.
The GaAs:Be, AlAs and GaMnAs layers were grown at 550~$^\circ$C,
530~$^\circ$C and 210~$^\circ$C, respectively.
We grew a Ga$_{0.94}$Mn$_{0.06}$As (25~nm) / AlAs (6~nm) / GaAs:Be QW (15~nm,
$p=1\times10^{19}$ cm$^{-3}$) / AlAs (6~nm) / GaAs:Be (100~nm,
$p=3\times10^{18}$ cm$^{-3}$) tunnel heterostructure for device Z.
The GaAs:Be electrode, AlAs, GaAs:Be QW and GaMnAs layers were grown at
570~$^\circ$C, 530~$^\circ$C, 400~$^\circ$C and 210~$^\circ$C, respectively.
The growth rates of GaAs, AlAs and GaMnAs were 500~nm/h.

After the growth, we annealed the samples in air at 180~$^\circ$C for 38 h for
devices A-C and 20 h for device Z to improve the crystallinity and
$T_\mathrm{C}$ of the GaMnAs layers \cite{Hayashi2001, Potashnik2001,
Potashnik2002, Yu2002, Edmonds2002}.
We estimated the $T_\mathrm{C}$ of the GaMnAs layers by measuring the magnetic
circular dichroism (MCD) on the samples and analysed the Arrott plots derived
from the MCD-$\mu_0|\mathbf{H}|$ curves at various temperatures.
The estimated $T_\mathrm{C}$ is 134~K for devices A-C and 116~K for device Z.

We fabricated tunnel diode devices on the wafer after growth.
In our process, we used chemical wet etching with an acid solution composed of
phosphoric acid, hydrogen peroxide and water.
For devices A-C,
we sank the wafer vertically into the etching liquid so that the thickness of
the GaMnAs layer changes from 0~nm to 25~nm in the same wafer.
Then, we made circular mesa devices with a diameter of 200 $\mu$m on the wafer
by chemical wet etching for devices A-C and Z.
We then coated a negative insulating resist on the
wafers for the passivation of the surfaces and opened
contact holes with a diameter of 180 $\mu$m on the mesa devices.
Then, we deposited Au on the wafers and fabricated contact pads on them.

We carried out the device process, especially the procedure from the surface
etching to the Au evaporation, very quickly to minimize the oxidation of the
surface GaMnAs layer. Indeed, the resonant levels systematically change with
the change in the thickness $d$ of the GaMnAs layer as shown in Supplementary
Figure 4. Also, similar results have been observed in tunnelling devices
with GaMnAs fabricated by similar methods \cite{Ohya2011, Muneta2016}.
Furthermore, the $\varphi$ dependence of $dI/dV$ around 0~V and in the positive
$V$ region is similar among samples A-C as shown in
Fig. \ref{fig:results-color-coded}a,d,g.
Therefore, extrinsic effects induced by the device process are negligible.

\noindent\textbf{Data procedure.}
We obtained the derivative of the $I$-$V$ characteristics numerically
using the Savitzky-Golay filter.
We used 7 data points to obtain the derivative at a single point.

In our fitting, we determined the fitting parameters by the modified
Levenberg-Marquardt least squares method.

\section*{Acknowledgements}
This work was partially supported by Grants-in-Aid for Scientific Research,
including Specially Promoted Research and the Project for Developing Innovation
Systems of MEXT.
Part of this work was carried out under the Cooperative Research Project
Program of RIEC, Tohoku University.
I.M. thanks the JSPS Research Fellowship Program for Young Scientists.

\section*{Author contributions}
Device fabrication and experiments: I.M., T.K.;
data analysis and theory: I.M.;
writing and project planning: I.M., S.O. and M.T.

\section*{Additional information}
The authors declare no competing financial interests.

\end{document}